\documentstyle{article}

\title {Numerical domain wall type solutions in $\phi^4$ theory}
\author {J. Karkowski 
	\thanks{Institute of Theoretical Physics, Jagellonian University,
	30-064 Krak\'{o}w, Reymonta 4, Poland}
	\ and
	Z. \'{S}wierczy\'{n}ski
	\thanks{Institute of Physics, Pedagogical University,
	 Podchor\c a\.{z}ych 2, 30-084 Krak\'{o}w, Poland}}

\begin{document}
\maketitle

\begin {abstract}
The well known domain wall type solutions are nowadays of great physical 
interest in classical field theory. These solutions can mostly be found only
approximatelly. Recently the Hilbert-Chapman-Enskog method was succesfully
applied to obtain this type solutions in $\phi^4$ theory. The goal of the 
present paper is to verify these perturbative results by numerical 
computations.
\end {abstract}

\vspace{0.5cm}
\indent
{\bf Introduction}
\vspace{.5cm}
\newline
\indent
During the last few years domain walls, strings and vortices have been
of great physical interest in classical and quantum field theory. These are
the solutions of highly non-linear partial differential equations with the
energy concentrated in a small region of the whole space-time. They play
an important role in cosmology, condensed matter and elementary
particles theory. Usually the exact form of these solutions can not be obtained
as the appropriate equations are too difficult to integrate. There are many
approximate methods which can help to solve these equations but they usually 
do not 
provide the quantitative estimation of errors. Thus they should be compared
with the numerical solutions obtained with help of the well tested algoritms.
Numerical methods can also verify many assumptions which must be made in 
perturbative calculations. In the present paper we consider the domain wall
type solutions in $\phi^4$ theory. We compare our numerical results with
the approximate solutions obtained in \cite{A0}. 
These approximate solutions were found with help of the so-called 
Hilbert-Chapman-Enskog method \cite {HCE}. It is the singular perturbative 
scheme \cite {LOM}
as the small perturbation parameter appears in the term with the highest order
derivatives. Thus from the mathematical point of view the problem is very
subtle and its numerical verification is especially valuable.
Let us stress that the paper \cite{A0}
is the part of the larger program \cite{AL1,A2,A3,AH4} devoted studying 
different approximate methods 
in classical field theory. Many technical problems have been examined in this
program so far and that is the second reason why we extend it 
on numerical analysis.   
\vspace{.5cm}
\newline
\indent
{\bf Approximate solutions}
\vspace{.5cm}
\newline
\indent
In this section we shortly describe the approximated domain wall type solutions
in $\phi^4$ theory proposed in \cite{A0}. The lagrangian density and the 
corresponding field equation have the form

$$
L = -{1\over 2}\eta_\mu{}_\nu\partial^\mu\Phi\partial^\nu\Phi
    - {\lambda \over 2}\left( \Phi^2 - {M^2 \over {4\lambda}} \right)^2 ,
\eqno(1)
$$

$$
\partial_\mu \partial^\mu\Phi
- 2\lambda\left( \Phi^2 - {M^2 \over {4\lambda}} \right)\Phi = 0 .
\eqno(2)
$$
Here $(\eta_\mu{}_\nu) = \hbox{diag}(-1,1,1,1)$ is the metric tensor
and $\lambda$ , $M$ are some positive constants. There exist two classical
vacuum solutions in this model, namely $\Phi = \pm M/2\sqrt{\lambda}$.
A domain wall type solution smoothly interpolates between these two
vacua. It is convenient to introduce in Minkowski space the special 
coordinate system $(u^0, u^1, u^2, s)$ co-moving with the domain wall
 \cite{A0,co}.
The three coordinates $u^0, u^1, u^2$ describe the world volume of the
membrane attached to the domain wall and $\xi=2s/M$ is the distance
from this three-dimensional hypersurface ( for details see \cite{A0} ).
Thus

$$
x^\mu = Y^\mu(u) + \xi n^\mu(u) ,
\eqno(3)
$$
where the functions $Y^\mu(u)$ describe the above mentioned hypersuface
and $n^\mu(u)$ is the normalised four-vector orthogonal to it.
In these new coordinates the equation of motion takes the form

$$
{2 \over M^2}{1 \over \sqrt{-G}}{\partial \over {\partial u^a}}
\left(\sqrt{-G}G^{ab}{\partial \phi \over {\partial u^b}}\right)
 + {1\over 2} {\partial^2 \phi \over {\partial s^2}}
 + {1\over {2h}}{\partial h \over {\partial s}}
    {\partial \phi \over {\partial s}}
 - (\phi^2 - 1)\phi = 0 ,
\eqno(4)
$$
where
 
$$
\Phi(x^\mu) = {M \over {2\sqrt{\lambda}}}\phi(u^a,s) ,
\eqno(5)
$$

$$
G_{ab} = N_{ac}g^{cd}N_{db}, N_{ac} = g_{ac} - \xi K_{ac},
\eqno(6)
$$

$$
\sqrt{-G} = \sqrt{-g} h, g = \det[g_{ab}], G = \det[G_{ab}].
\eqno(7)
$$
The induced metric $g_{ab}$ and the external curvature $K_{ac}$ 
are defined by

$$
K_{ab} = n_\mu \partial_a \partial_b Y^\mu ,
g_{ab} = \partial_a Y^\mu \partial_b Y_\mu .
\eqno(8)
$$
The Hilbert-Chapman-Enskog method has been applied in \cite{A0} to obtain
the perturbative solution of the equation (4). According to this method
the following form is assumed for the approximate solution of the equation (4)

$$
\phi = \phi^{(0)} + {1\over M}\phi^{(1)} + {1\over M^2}\phi^{(2)} + \ldots \ .
\eqno(9)
$$
In the present paper this approximation is verified numerically to give very
precise results. However in general the problem is a little bit more 
complicated
as the equation (4) involves the small perturbation parameter in the term 
with the highest order derivative. Usually in such case one can expect
the solution to involve some extra non-analitic terms which of course are 
absent
in the formula (9). This problem is discussed mainly from the physical point 
of view
in \cite{A0}. The non-analitic terms can really exist in the general solution
but they are only radiative corrections to the base domain-wall configuration.
These terms can be completely removed from the solution by the appropriate
choice of the initial conditions. This is the case we are especially interested
in as we would like to extract the pure domain-wall type solution without
radiation. In practice it is very difficult to find the desired initial 
conditions. However even if the radiative corrections are present they quite 
rapidly decrease with time as the whole solution tends to the stable
domain-wall
type configuration. Thus these possible extra terms can practically be 
neglected. This is the reason why the Hilbert-Enskog-Chapman method could be
safely applied in \cite{A0}. We will return to this 
problem in our conclusions. 
Assuming that for $u^0=0$ the surface $\phi=0$ coincides with 
the membrane $s=0$ the result looks as follows
$$
\phi = \tanh s + {1\over M}{C\over {\cosh^2s}}
$$
$$ 
- {1\over {M^2}}\left[
 C^2 \left({{6f_1(s)}\over \cosh^2s} +
 {{\phi_2^{(2)}(s)}\over \cosh^6s}\right) +
 4K_a^bK_b^a\left({{f_1(s)}\over \cosh^2s} + f_3(s){\phi_2^{(2)}(s)}\right) 
\right] + ... \ ,
\eqno(10)
$$
where

$$
\phi_2^{(2)} = {1\over 8}\sinh(2s) + {3\over 8}\tanh s
  + {3\over 8}{s\over {\cosh^2 s}} ,
\eqno(11)
$$

$$
f_1(s)=\int_0^s dx{{\sinh x}\over{\cosh^5x}}\phi_2^{(2)}(x) ,
\eqno(12)
$$

$$
f_2(s)=\int_0^s dx{x\over{\cosh^2x}}\phi_2^{(2)}(x) ,
\eqno(13)
$$

$$
f_3(s)=- \int_0^s dx{x\over{\cosh^4x}}
\eqno(14)
$$
and the function $C(u^a)$ satisfies the equation

$$
\Box^{(3)}C + \left({{\pi^4}\over 4} - 1\right)K_a^bK_b^a C
 + {9\over 35}C^3 = \left({{\pi^2}\over 6} - 1\right)K_c^aK_a^bK_b^c ,
\eqno(15)
$$
with initial conditions 
$$
C(u^0=0)={\partial C \over \partial u^0}(u^0=0)=0 .
\eqno(16)
$$

Here the symbol $\Box^{(3)}$ denotes the three-dimensional d'Alembertian 
on the membrane surface. The function $C(u^a)$ determines the position 
of the surface where the field $\phi$ vanishes. This position is given 
by the formula

$$
s = -{C\over M} + O(M^{-3}) .
\eqno(17)
$$
\vspace{.5cm}
\newline
\indent
{\bf Numerical results}
\vspace{.5cm}
\newline
\indent
We have examined in detail two special cases of the general result (10),
namely the cylindrical and the spherical domain walls. In the cylindrical
case the function $C(u^a)$ and the invariant $K_c^aK_a^bK_b^c$ vanish.
The other invariant $K_b^aK_a^b$ takes the simple form

$$
K_a^bK_b^a=2 {r_0^2 \over r^4} ,
\eqno(18)
$$
where

$$
r(u^0) = r_0 \cos {u^0 \over r_0}
\eqno(19)
$$
and $r_0$ is the initial position of the domain wall. Let us stress that 
in this case the position of the surface where the field $\phi$ vanishes 
coincides with the membrane attached to the domain wall, at least up to 
the second order in $1/M$. The equation of 
motion (4) for the cylidrical domain wall reads as

$$
{\partial \over {\partial u^0}} \left( 
 { {r+\xi{r_0 \over r}} \over {{r \over r_0} - {\xi \over r}} }
   {{\partial \phi} \over {\partial {u^0}}}  \right) +
 { \left( r+\xi{r \over {r_0}} \right)  
    \left( {{r \over {r_0}} - {\xi \over r}} \right) } 
$$
$$
\times \left(
{{\partial^2 \phi} \over {\partial \xi^2}} 
- 2 {{\xi r_0^2} \over {r^4 - {\xi}^2 r_0^2}} {{\partial \phi}\over 
{\partial \xi}}
- {M^2 \over 2} \left( \phi^2 - 1 \right) \phi 
\right) = 0 .
\eqno(20)
$$
The spherical case is a little bit more complicated because both invariants
$K_b^aK_a^b$ and $K_c^aK_a^bK_b^c$ are different from zero

$$
K_a^bK_b^a=6 {r_0^4 \over r^6} ,
\eqno(21)
$$

$$
K_c^aK_a^bK_b^c = 6 {r_0^6 \over r^9} ,
\eqno(22)
$$
where

$$
r(u^0) = r_0 \hbox{cn} \left(\sqrt{2}{{u^0}\over {r_0}}, {1\over 2}\right)
\eqno(23) .
$$
As a consequence the function $C(u^a)$ does not vanish and should be
obtained numerically. Thus in this case the distance between the above 
mentioned surfaces is different from zero. The appropriate equation of 
motion for the spherical domain wall takes the form

$$
{\partial \over {\partial u^0}} \left( 
 { {r+\xi {r_0^2 \over r^2}} \over {{r^2 \over {r_0^2}} - 2 {\xi \over r}} }
   {{\partial \phi} \over {\partial {u^0}}}  \right) +
 { \left( r+\xi{r_0^2 \over {r^2}} \right)  
    \left( {{r^2 \over {r_0^2}} - 2 {\xi \over r}} \right) }
$$
$$
\times \left(
{{\partial^2 \phi} \over {\partial \xi^2}} 
- 6 {{\xi r_0^4} \over {r^6 -\xi r^3 {r_0}^2 -2 \xi^2 r_0^4}} 
{{\partial \phi}\over {\partial \xi}}
- {M^2 \over 2} \left( \phi^2 - 1 \right) \phi 
\right) = 0 .
\eqno(24)
$$
We have solved the equations (20) and (24) numerically for 
$-0.5 \leq \xi\leq 0.5$ and $0 \leq u^0 \leq 1$ with the initial data 
$\phi|_{u^0=0}, \partial \phi /\partial u^0|_{u^0=0}$ computed 
from formula (10) with $K_b^aK_a^b$ given by (18) or (21) respectively.
Let us note that the maximal value of $u^0$ can not be too big because of
possible singularities in coordinate system \cite{AL1}.
We have introduced a grid of 8000 points and the partial derivatives 
with respect to $s$ have been replaced with

$$
{{\partial \phi}\over {\partial s}} \approx 
 { {\phi_{n+1} - \phi_{n-1}}\over {2h} } ,
{{\partial \phi^2}\over {\partial^2 s}} \approx 
 { {\phi_{n+1} - 2\phi_n + \phi_{n-1}}\over {h^2} } .
\eqno(25)
$$ 
The system of ordinary differential equations obtained this way has been 
integrated by the Runge-Kutta method \cite{NR}. The numerical computations
in both
cases e.g. cylindrical and spherical have been performed for $r_0=2.5$ 
and three different values of the parameter $M=10,20,40$.
We have compared the numerical
functions $\phi_{num}(u^0=1.0)$ and $\partial\phi_{num}/\partial u^0(u^0=1.0)$
with their approximate counterparts $\phi_{ap}(u^0=1.0)$ and
$\partial\phi_{ap}/\partial u^0(u^0=1.0)$
obtained from formula (10). The motion of the surface $\phi=0$ has
been also computed and compared with the approximate result (17).
Figures (1)-(4) show the differences $\Delta\phi=\phi_{num}-\phi_{ap}$
and $\Delta\phi_{u^0}=\partial\phi_{num}/{\partial u^0}-
\partial\phi_{ap}/{\partial u^0}$ in the case of the cylindrical domain wall.
We do not present the functions $\phi_{num}$ themselves because the interesting
higher order corrections are 
very small compared to the $u^0$ independent term $\phi^{(0)}= \tanh(s)$.
Instead the difference $\phi_{num}-\tanh(s)$ is presented in Fig(5) for
$M=20$. The corresponding derivative $\partial \phi / \partial u^0$ is shown
in Fig(6). 
The motion of the core $\phi_{num}=0$ is plotted in figures (13) - (15). 
Let us recall
that in this case $C\equiv0$ and the surface $\phi_{ap}=0$ coincides with the
membrane $s=0$. The differences $\Delta\phi$ and $\Delta\phi_{u^0}$ in the 
spherical case are presented in figures (7) - (10).
Figures (11) and (12) are the spherical counterparts of (5) and (6).
The motion of the core $\phi_{num}=0$
and the correspondig results
computed from the formula (17) for $\phi_{ap}=0$ 
are plotted in figures (16) - (18).
\vspace{.5cm}
\newline
\indent
{\bf Conclusions}
\vspace{.25cm}
\newline
\indent
The aim of our paper was to verify the approximate solutions proposed in 
\cite{A0}
by the numerical computations. Our investigations prove that the coincidence
of the numerical results and the approximate ones is quite good.
The existing differences could be affected by radiative corrections absent
in the perturbative scheme. 
In numerical calculations these corrections can not be completely removed.
However our results show that these possible terms are small and can be 
neglected in our considerations as the perturbative solution is only known 
up to the second order term in $ 1/M $. It would be interesting to compare
these radiative corrections with the higher order terms of the perturbative
scheme. This is the argument for further studies of these 
important questions.
As it should be expected the differences $\phi_{num}-\phi_{ap}$ and
$\partial \phi_{num} / \partial u^0 - \partial \phi_{ap}/\partial u^0$
decrease rather rapidly when the models with thinner walls (bigger $M$)
are considered. Let us also stress that the numerical and the approximated
motion of the domain wall almost coincide for our biggest value
of the parameter $M$. Thus the approximation proposed in \cite{A0} seems to be
an efficient tool to examine important details of some peculiar 
solutions in classical field theory like domain walls or vortices.
In general our paper proves that the progress in the practical dealing 
with non-linear partial differential
equations can be achieved when one combines different clever approaches
such as the Hilbert-Enskog-Chapman perturbative method and the suitable 
choice of the coordinate system.

\pagebreak

\end{document}